\documentclass[12pt, a4paper, reqno]{amsart}           

\usepackage[round, comma, authoryear, sort]{natbib}

\usepackage[
    colorlinks = true, 
    final = true, bookmarks = true, bookmarksnumbered = true, 
    bookmarksopen = true, 
    setpagesize = false, 
    pdfpagelayout = SinglePage, 
    pdftitle = {{
        A branch-and-price approach with MILP formulation 
        to modularity density maximization 
        on graphs
    }}, 
    pdfauthor = {{Keisuke Sato and Yoichi Izunaga}}, 
    pdfkeywords = {{
        Graph clustering, 
        Modularity density, 
        Set partitioning, 
        Branch-and-price, 
        Set-packing relaxation, 
        Multiple cutting planes.
    }}, 
]{hyperref}

\usepackage{graphicx}

\usepackage{tikz}

\usepackage{amsmath,amssymb,amsthm}
\usepackage{amsaddr}

\usepackage{enumitem}

\usepackage[T1]{fontenc}
\usepackage{lmodern}

\usepackage{exscale}

\usepackage{algorithm}
\usepackage[noend]{algpseudocode}

\renewcommand{\emptyset}{\varnothing}
\renewcommand{\Re}{\mathbb{R}}

\newcounter{mpproblem}[section]

\makeatletter
\newenvironment{mpproblem}[1]%
    {%
        \protected@edef\@currentlabelname{#1}%
        \par\vspace{\baselineskip}\noindent%
        \ifx#1\empty %
        \else \refstepcounter{mpproblem}$($#1$)$ %
        \fi%
        \hfill%
        $\left|%
        \hfill%
        \hspace{0.00\textwidth}%
        \@fleqntrue\@mathmargin\parindent%
        \begin{minipage}{0.86\textwidth}%
        \vspace{-1.0\baselineskip}%
    }%
    {%
        \end{minipage}%
        \@fleqnfalse%
        \right.$%
        \par\vspace{\baselineskip}\noindent%
        \ignorespacesafterend%
    }%
\makeatother
\newcommand{\mpprobref}[1]{$($\nameref{#1}$)$}

\newenvironment{mpproblem*}%
    {%
        \begin{mpproblem}{}%
    }%
    {%
        \end{mpproblem}%
        \ignorespacesafterend%
    }

\title[Branch-and-price with MILP to modularity density]%
{
    A branch-and-price approach with MILP formulation 
    to modularity density maximization 
    on graphs
}

\author{Keisuke Sato}
\address{
    Signalling and Transport Information Technology Division, 
    Railway Technical Research Institute. 
    2-8-38 Hikari-cho, Kokubunji-shi, Tokyo 185-8540, Japan
}
\email{(Keisuke Sato) sato.keisuke.49@rtri.or.jp}

\author{Yoichi Izunaga}
\address{
    Information Systems Research Division, 
    The Institute of Behavioral Sciences. 
    2-9 Ichigayahonmura-cho, Shinjyuku-ku, Tokyo 162-0845, Japan
}
\email{(Yoichi Izunaga) yizunaga@ibs.or.jp}

\date{\today}

\thanks{
\textbf{
    The earlier version (v2) of this paper 
    is available at \url{https://arxiv.org/abs/1705.02961v2}, 
    which was published on May~10, 2017. 
}
}

\begin{document}
\pdfbookmark[1]{Title}{Title}
\begin{abstract}
\hspace*{1mm}
For clustering of an undirected graph, 
this paper presents an exact algorithm 
for the maximization of modularity density, 
a more complicated criterion 
to overcome drawbacks of the well-known modularity. 
The problem can be interpreted as the set-partitioning problem, 
which reminds us of its integer linear programming~(ILP) formulation. 
We provide a branch-and-price framework for solving this ILP, 
or column generation combined with branch-and-bound. 
Above all, 
we formulate the column generation subproblem to be solved repeatedly 
as a simpler mixed integer linear programming~(MILP) problem. 
Acceleration techniques called 
the set-packing relaxation and the multiple-cutting-planes-at-a-time 
combined with the MILP formulation 
enable us to optimize the modularity density 
for famous test instances 
including ones with over 100 vertices in around four minutes by a PC. 
Our solution method is deterministic 
and the computation time is not affected 
by any stochastic behavior. 
For one of them, 
column generation at the root node of the branch-and-bound tree 
provides a fractional upper bound solution 
and our algorithm finds an integral optimal solution after branching. 
\end{abstract}

\maketitle


\section{Introduction}
\label{sec:Introduction}
Identifying communities in graphs is a very important task in data analysis, 
and has a wide range of applications in diverse fields 
such as social networks, the Web, biology and bioinformatics. 
Roughly speaking, a community is a subset of a graph which are tightly connected internally 
while loosely connected externally. 
Numerous approaches to community detection have been proposed so far, 
most of which aim to optimize a certain objective function defined on a graph. 

Triggered by the seminal work by \citet{Newman.2004} in the literature of the community detection, 
maximizing the {\it modularity} function has extensively been studied. 
Let $G := (V, E)$ be an undirected graph 
with the set $V$ of $n$ vertices where $n \geq 2$ 
and the set $E$ of $m$ edges. 
The degree of vertex $i \in V$ is denoted by $d_i$. 
We say that $\Pi := \{ C_1, \ldots, C_k \}$ 
for some positive integer $k$ 
is a \emph{partition} of $V$ 
if $V = \cup_{p = 1}^{k} C_p$, $C_p \not= \emptyset$ for any $p$ 
and $C_p \cap C_{p^\prime} = \emptyset$ for any distinct pair $p$ and $p^\prime$ hold. 
Each member $C_p$ of a partition is called a \emph{community}. 
The set of edges that have one end-vertex in community $C$ 
and the other end-vertex in community $C^\prime$ is denoted by $E(C, C^\prime)$. 
When $C = C^\prime$, 
we abbreviate $E(C, C^\prime)$ to $E(C)$ for the sake of simplicity. 
Then the modularity, denoted by $Q(\Pi)$ for partition $\Pi$ of $V$, 
is defined as 
%
\begin{equation*}
Q(\Pi) := 
\sum_{C \in \Pi}
\left(
        \frac{|E(C)|}{m} - \left( \frac{\sum_{i \in C} d_i}{2m} \right)^2 
\right)
\end{equation*}
%
where $|\cdot|$ is the cardinality of the corresponding set.

The modularity maximization is now one of the central subjects in this field, 
while it receives serious criticism from mainly two viewpoints: 
\emph{degeneracy} (\citet{Good.2010}) and \emph{resolution limit} (\citet{Fortunato.2007}). 
Degeneracy means presence of several partitions with high modularity 
which makes it difficult to find a global optimal partition. 
Resolution limit refers to sensitivity of modularity to the total number of edges in a graph, 
which leaves small communities not identified and hidden inside larger ones. 
Even in a schematic case where a graph consists of multiple replicas of an identical clique 
which are connected by a single edge, 
\citet{Fortunato.2007} showed that maximizing the criterion 
results in regarding two or more cliques connected as a community 
when the number of cliques in the graph is larger than the square root of the number of edges.
This narrows an application area of the modularity maximization 
since most of real networks may contain tightly connected groups with different scales. 

To avoid the resolution limit issue, 
\citet{Li.2008}~\footnote{
Based on comments on this paper by \citet{Costa.2014}, 
errata by \citet{Li.2015} were released. 
}
proposed a new function, called {\it modularity density},
and their theoretical analysis 
with respect to maximizing the function leads to detecting communities 
with different scales. 
The modularity density, denoted by $D(\Pi)$ for partition $\Pi$, is defined as 
%
\begin{equation*}
D(\Pi) := 
\sum_{C \in \Pi} 
\left(
        \frac{  2|E(C)| - \sum_{C^\prime \in \Pi \setminus \{ C \}} |E(C, C^\prime)|}{|C|} 
\right).
\label{eqn:MD_Def}
\end{equation*}
%
We refer to each term of the outer summation in $Q(\Pi)$ or $D(\Pi)$ 
as the \emph{contribution} of the community 
to the modularity or the modularity density. 

Since this function takes account of the number of vertices in each community,
the modularity density maximization is straightforwardly formulated 
as a binary nonlinear fractional programming problem. 
This feature indicates 
that development of any exact solution method 
for the modularity density maximization 
seems to be more challenging than that for the modularity maximization, 
which has promoted development of heuristic algorithms. 
In fact, \citet{Li.2008} fixed the number of communities 
and solved the continuous relaxation problem. 
Although \citet{Karimi.2015} presented an improved formulation 
that does not require the number of communities to be known, 
it is still a binary nonlinear fractional programming problem. 
To date, several metaheuristic approaches have been developed: 
ones based on a genetic algorithm by \citet{Liu.2010}, 
a memetic algorithm with simulated annealing 
in its local search phase by \citet{Gong.2012} 
and biological operations by \citet{Karimi.2015}. 
\citet{Costa.2016} proposed hierarchical divisive heuristics 
based on repetitive resolutions 
of an integer linear programming~(ILP) problem 
or a mixed integer linear programming~(MILP) problem 
to split a community into two. 
\citet{Santiago.2017} presented seven scalable heuristic methods, 
and compared them with the metaheuristic algorithms mentioned above 
as well as the heuristics by \citet{Costa.2016}. 
\citet{Izunaga.2016} formulated the problem 
as a variant of a semidefinite programming problem called 0-1SDP. 
Their reformulation has the advantage 
that it does not require the number of communities in advance, 
while any method to exactly optimize 0-1SDP has yet been unavailable. 
Instead, they solved an ordinary semidefinite programming 
relaxation problem to obtain an upper bound solution 
and created a feasible solution from it by dynamic programming. 

On the other hand, 
there are a few approaches to exactly maximize the modularity density. 
The exact formulation 
proposed by \citet{Li.2008}, \citet{Karimi.2015} or \citet{Izunaga.2016} 
has not yet been solved to optimality due to its nonlinearity. 
\citet{Costa.2015} presented several MILP reformulations, 
which enables us an application of general-purpose optimization solvers to the problem. 
In the reformulations, however, the number of communities must be fixed in advance. 
They reported a result that 
their best MILP formulation gave optimal solutions of instances with at most 40 vertices. 
The models require the upper bound value of the contribution of a community as input, 
which was calculated by solving a binary nonlinear fractional programming problem 
in the paper. 
\citet{Costa.InPress} discussed MILP reformulations of the upper bound calculation, 
providing the whole modularity density maximization process 
completely expressed as MILP problems. 
\citet{Izunaga.2016} calculated the upper bound 
in their numerical experiments for comparison 
by the parametric algorithm by \citet{Dinkelbach.1967} 
in which a series of ILPs was solved. 

Very recently, and independently of our work, 
\citet{Santiago.InPress} have considered the clustering problem 
as the set-partitioning problem 
and have presented its ILP formulation 
(refer, for instance, to \citet{Nemhauser.1999} 
on the set-partitioning and related problems as well as their ILP formulations). 
They have solved the problem by column generation 
(refer, for instance, to \citet{Desrosiers.2005} on column generation), 
in which framework 
an initial set of columns is given by heuristics 
that has stochastic behavior. 
The column generation subproblem 
is also solved by different stochastic heuristics. 
When no column is found by the latter heuristics, 
the subproblem is formulated as an integer quadratic programming~(IQP) problem 
and is solved to optimality 
to decide whether the linear programming~(LP) relaxation 
of the set-partitioning problem 
is optimal or not even though the LP has a limited set of columns. 
Although they have reported optimal solutions for instances having 62 and 105 vertices, 
the computation time has varied considerably for each trial 
due to the stochastic nature of the two heuristics. 
For several trials, these instances have not been solved in ten hours. 
Another remark should be made 
that they have only solved instances whose LP optimal solution is integral 
and have not presented a detailed procedure 
for a case where the LP optimal solution is fractional. 
Hence, their approach may not 
provide an optimal solution for a particular unsolved instance. 

In this paper, independently of the work by \citet{Santiago.InPress}, 
we regard the modularity density maximization 
as the set-partitioning problem 
and present its ILP formulation, 
which enables us to devise an efficient algorithm to provide an optimal solution 
for the modularity density maximization. 
To be specific, we develop an algorithm based on a branch-and-price framework, 
i.e., column generation in a branch-and-bound framework, 
to truly, exactly optimize the modularity density function value 
(refer to \citet{Barnhart.1998} as well as \citet{Desrosiers.2005} 
on branch-and-price). 
We also incorporate two existing techniques into the algorithm: 
the set-packing relaxation proposed by 
\citet{Sato.2012}~\footnote{%
    More accurately, 
    they presented the set-covering relaxation 
    since they aimed to solve a minimization problem. 
}, 
which was originally applied to a set-partitioning-based scheduling problem, 
and the multiple-cutting-planes-at-a-time by \citet{Izunaga.2017}, 
which was originally done to the modularity maximization, 
to accelerate the column generation process within the algorithm. 
The former substitutes the set-partitioning constraint 
of the LP relaxation problem with the set-packing constraint 
for all the vertices, 
and dynamically restoring it for a necessary subset of the vertices 
in the column generation process. 
We expect that the contribution of the majority of communities 
detected as an optimal solution will take a positive value, 
and therefore that the set-packing constraint will suffice 
for a large part of the vertices. 
The latter can provide us with two or more columns 
that have no common element in each column generation phase, 
and therefore we expect that these columns will coexist 
in a good feasible solution of the original or the LP relaxation of 
the set-partitioning/set-packing problem. 

Our contributions in this paper can be summarized 
as follows: 
\begin{enumerate}
\item We give a branch-and-price framework 
for the exact modularity density maximization problem 
expressed as the ILP formulation of the set-partitioning problem. 
For Protein~p53 test instance having 104 vertices, 
we show that column generation at the root node of the branch-and-bound tree 
provides a fractional upper bound solution 
and that our algorithm finds an integral optimal solution after branching. 

\item We formulate the column generation subproblem 
to be solved repeatedly 
as a simpler MILP problem than the quite recently proposed IQP problem. 
This formulation lets us provide another complete MILP framework 
for the whole modularity density maximization process. 

\item The set-packing relaxation and the multiple-cutting-planes-at-a-time 
acceleration techniques 
combined with the MILP formulation 
of the column generation subproblem 
enable us to optimize the modularity density 
for famous test instances 
including Graph, Dolphins, Les~Mis\'{e}rables and A00~main 
in addition to Protein~p53, which have not yet been solved. 
Instances with over 100 vertices are solved with a proof of optimality 
in around four minutes by a PC. 
Our solution method is deterministic 
and the computation time is not affected 
by any stochastic behavior. 
\end{enumerate}

The rest of the paper is organized as follows: 
in Section~\ref{sec:Set-partitioning}, 
we present the set-partitioning formulation 
of the modularity density maximization 
and column generation for that, 
referring to the recently proposed IQP formulation 
of the column generation subproblem by \citet{Santiago.InPress}. 
In Section~\ref{sec:BranchAndPrice}, 
we propose a solution framework based on branch-and-price. 
It includes a MILP formulation of the column generation subproblem 
as well as the set-covering relaxation and the multiple-cutting-planes-at-a-time 
acceleration techniques. 
In Section~\ref{sec:Results}, 
we report numerical experiments on the proposed framework. 
Finally, conclusions and future work are presented 
in Section~\ref{sec:Conclusions}. 

\section{Set-partitioning and column generation at root node}
\label{sec:Set-partitioning}
\subsection{Set-partitioning ILP formulation}
\label{sec:Set-partitioning_SetPartition}
\mbox{ }

Any feasible solution to the modularity density maximization 
as well as the modularity maximization 
is a partition of the vertex set. 
Hence, as it was done to the modularity maximization 
by \citet{Aloise.2010}, 
we can regard the modularity density maximization 
as the set-partitioning problem. 
The problem is widely formulated as an ILP problem. 

The set of all possible communities is $2^V \setminus \{ \emptyset \}$, 
and we let it be $\mathcal{C}$. 
Any possible community $C \in \mathcal{C}$ 
satisfies $\emptyset \subsetneq C \subseteq V$, 
i.e., $C$ consists of some members of $V$. 
Given $C \in \mathcal{C}$, 
its contribution $f_C$ to the modularity density is calculated by 
%
\begin{equation*}
f_C = \frac{ 4|E(C)| - \sum_{i \in C} d_i}{|C|}. 
\end{equation*}
%
Let constant $a_{iC}$ be one 
if vertex $i \in V$ is in possible community $C \in \mathcal{C}$ 
and be zero otherwise. 
Also, we let $u_{C}$ be a binary variable 
indicating whether $C \in \mathcal{C}$ is selected for a community or not. 
Then the set partitioning formulation~\mpprobref{mpprob:P} 
of the modularity density maximization 
is as follows: 
\begin{mpproblem}{P}
\label{mpprob:P}
\begin{alignat*}{3}
 &\text{max.} & \quad \sum_{C \in \mathcal{C}} f_C u_C      &
     & \quad &                           \\
 &\text{s.t.} & \quad \sum_{C \in \mathcal{C}} a_{iC} u_{C} &= 1 
     & \quad &\forall i \in V            \\
 &            & \quad u_C                                   &\in \{0 ,1\} 
     & \quad &\forall C \in \mathcal{C}. 
\end{alignat*}
\end{mpproblem}
The main advantage of this approach 
is that we do not need to give the optimal number of communities 
in advance.

\subsection{Subproblem as IQP in column generation at root node}
\label{sec:Set-partitioning_IQP}
\mbox{ }

It is natural to rely on column generation
to solve \mpprobref{mpprob:P} 
since $|\mathcal{C}|$, which is equivalent to the number of variables, 
becomes extremely large 
as the number of vertices $n$ gets larger. 
This makes the problem intractable. 
Here let us introduce what we call column generation ``at the root node,'' 
which has also been presented by \citet{Santiago.InPress} quite recently. 
In the column generation process, 
\mpprobref{mpprob:P} is called the master problem, 
and a restricted master problem of \mpprobref{mpprob:P} 
is commonly given by substituting subset of $\mathcal{C}$ 
and continuously relaxed $u_{\mathcal{C}}$. 
Let $\ell \in \{1, 2, \ldots, \}$ 
be an iteration counter of the column generation, 
and the restricted master problem at the root node~\mpprobref{mpprob:RP-Root} 
is given by 
\begin{mpproblem}{RP$_{(\ell)}$}
\label{mpprob:RP-Root}
\begin{alignat*}{3}
 &\text{max.} & \quad \sum_{C \in \mathcal{C}_{(\ell)}} f_C u_C      &
     & \quad &                                    \\
 &\text{s.t.} & \quad \sum_{C \in \mathcal{C}_{(\ell)}} a_{iC} u_{C} &= 1 
     & \quad &\forall i \in V                     \\
 &            & \quad u_C                                            &\geq 0 
     & \quad &\forall C \in \mathcal{C}_{(\ell)} 
\end{alignat*}
\end{mpproblem}
where $\mathcal{C}_{(\ell)} \subseteq \mathcal{C}$ for each $\ell$. 
Possible community $C \in \mathcal{C}$ is called a column in this context. 
Let \mpprobref{mpprob:RD-Root} be the dual of~\mpprobref{mpprob:RP-Root} 
and $\lambda_i$ for vertex $i \in V$ be the dual variable. 
Then the problem is written as 
\begin{mpproblem}{RD$_{(\ell)}$}
\label{mpprob:RD-Root}
\begin{alignat*}{3}
 &\text{min.} & \quad \sum_{i \in V} \lambda_i        &
     & \quad &                                     \\
 &\text{s.t.} & \quad \sum_{i \in V} a_{iC} \lambda_i &\geq f_C 
     & \quad &\forall C \in \mathcal{C}_{(\ell)} \\
 &            & \quad \lambda_i                       &\in \Re 
     & \quad &\forall i \in V.                     
\end{alignat*}
\end{mpproblem}
Note that \citet{Santiago.InPress} have generated 
30 columns to form the initial column set $\mathcal{C}_{(1)}$ 
by heuristics that has stochastic behavior. 

In the column generation process, 
we solve \mpprobref{mpprob:RP-Root} or \mpprobref{mpprob:RD-Root} 
for each $\ell$, 
and try to generate column $\widehat{C} \in \mathcal{C} \setminus \mathcal{C}_{(\ell)}$ 
which has the possibility 
of improving the objective value of \mpprobref{mpprob:RP-Root} 
or equivalently cutting the optimal solution to \mpprobref{mpprob:RD-Root} 
by adding $\widehat{C}$ to $\mathcal{C}_{(\ell)}$. 
We define $\lambda^\ast_{i,(\ell)}$ for $i \in V$ 
as the dual price of the constraint for $i$ 
at an optimal solution to \mpprobref{mpprob:RP-Root} 
or as an optimal solution to \mpprobref{mpprob:RD-Root}. 
Let us focus on the dual problem, 
and column $\widehat{C} \in \mathcal{C} \setminus \mathcal{C}_{(\ell)}$ 
to cut the optimal solution 
must satisfy the following inequality: 
%
\begin{equation*}
\sum_{i \in V} a_{i \widehat{C}} \lambda^\ast_{i,(\ell)} < f_{\widehat{C}}. 
\end{equation*}
%
Now let us introduce binary variable $x_{i}$ 
which takes one if $i$ belongs to a column to be added and zero otherwise. 
The vector of $x_{i}$'s is denoted by \mbox{\boldmath $x$}. 
Then the search for such a column called the column generation subproblem 
at the root node~\mpprobref{mpprob:S-Root} is given by 
\begin{mpproblem}{S$_{(\ell)}$}
\label{mpprob:S-Root}
\begin{alignat*}{2}
 &\text{find}      & \quad &\mbox{\boldmath $x$} \in \{0, 1\}^V \setminus \{ 0^V \} \\
 &\text{such that} & \quad &\frac{  4 \sum_{\{i,j\} \in E} x_{i} x_{j} 
                                  - \sum_{i \in V} d_i x_{i}}
                                 {\sum_{i \in V} x_{i}} 
                            - \sum_{i \in V} \lambda^\ast_{i,(\ell)} x_{i} > 0.     
\end{alignat*}
\end{mpproblem}
To find any solution to this problem, 
\citet{Santiago.InPress} have presented different stochastic heuristics. 
In case of the failure, 
they have given the following equation: 
\begin{eqnarray*}
{} & & \frac{  4 \sum_{\{i,j\} \in E} x_{i} x_{j} 
             - \sum_{i \in V} d_i x_{i}}
            {\sum_{i \in V} x_{i}} 
       - \sum_{i \in V} \lambda^\ast_{i,(\ell)} x_{i} \\
{} &=& \frac{  4 \sum_{\{i,j\} \in E} x_{i} x_{j} 
             - \sum_{i \in V} d_i x_{i}
             - \sum_{i \in V} \sum_{i^\prime \in V} \lambda^\ast_{i,(\ell)} x_{i} x_{i^\prime}}
            {\sum_{i \in V} x_{i}}, 
\end{eqnarray*}
and have focused on its numerator. 
They have introduced variable $w_{ij}$ for each $(i,j) \in E$ which 
takes one if $x_i = x_j = 1$ and zero otherwise, 
and have defined the exact formulation of the subproblem at the root node 
as the following IQP: 
\begin{mpproblem}{S$^{\textsf{IQP}}_{(\ell)}$}
\label{mpprob:SS-IQP}
\begin{alignat*}{3}
 &\text{max.}        & \quad   4 \sum_{ \{ i,j \} \in E} w_{ij} 
                             - \sum_{i \in V} d_i x_{i} 
                             - \sum_{i \in V} \sum_{i^\prime \in V} \lambda^\ast_{i,(\ell)} x_{i} x_{i^\prime}
                                                  &               
     & \quad &                         
\end{alignat*}
\begin{alignat*}{3}
 &\text{s.t.} \;\;\; & \quad w_{ij}               &\leq x_{i}     
     & \quad &\forall \{ i, j\} \in E  \\
 &                   & \quad w_{ij}               &\leq x_{j}     
     & \quad &\forall \{ i, j\} \in E  \\
 &                   & \quad x_{i}                &\in \{ 0, 1 \} 
     & \quad &\forall i \in V          \\
 &                   & \quad w_{ij}               &\in \{ 0, 1 \} 
     & \quad &\forall \{ i, j\} \in E. 
\end{alignat*}
\end{mpproblem}
They have called the problem (AP-II), and have solved it to optimality. 
Note that \mpprobref{mpprob:SS-IQP} is a maximization problem 
and therefore that $w_{ij} = 1$ holds when $x_i = x_j = 1$ 
at an optimal solution. 
Although its objective function is nonconvex, 
it can be cast as an equivalent convex programming problem 
(refer, for instance, to \citet{Billionnet.2007}). 
Several solvers automatically perform the conversion, 
hence can handle \mpprobref{mpprob:SS-IQP}. 

If any \mbox{\boldmath $x$} is found 
that satisfies the condition of \mpprobref{mpprob:S-Root}, 
set $\widehat{C} := \{ i \in V \mid x_i = 1 \}$ 
is identified as a new, generated column. 
It is added to $\mathcal{C}_{(\ell)}$, 
which forms $\mathcal{C}_{(\ell+1)}$. 
Then \mpprobref{mpprob:RP-Root} or \mpprobref{mpprob:RD-Root} for $\ell+1$ is solved. 
In the former problem, the variable $u_{\widehat{C}}$ 
as well as the column vector $[a_{i \widehat{C}}]^\top_{i \in V}$ 
has been generated. 
In the latter problem, 
the constraint $\sum_{i \in V} a_{i\widehat{C}} \lambda_i \geq f_{\widehat{C}}$, 
which can be regarded as a cutting plane, has been added. 
If there is no such \mbox{\boldmath $x$}, 
we have an optimal solution to the LP relaxation of \mpprobref{mpprob:P}. 
For each of the instances 
which have been solved by \citet{Santiago.InPress}, 
the solution is integral, thereby indicating 
that it is an optimal solution to the modularity density maximization. 

\section{Branch-and-price with acceleration techniques}
\label{sec:BranchAndPrice}
\subsection{Branch-and-price framework}
\label{sec:BranchAndPrice_BranchAndPrice}
\mbox{ }

Let us consider the possibility, 
for a particular unsolved instance, 
that the column generation at the root node 
presented in the previous section 
provides a fractional solution. 
Then the solution is of course infeasible 
in terms of the original ILP problem, 
or an integral solution 
obtained by solving the ILP set-partitioning problem 
given the set of columns which have been generated until then 
has not been proven to be exactly optimal in general. 
This possibility necessitates a branch-and-price framework, 
or column generation combined with branch-and-bound. 

We follow the standard ``identical restrictions on subsets'' branching rule 
for the set-partitioning problem by \citet{Barnhart.1998}, 
which dates back to \citet{Ryan.1981}. 
Let $b \in \{ 0, 1, \ldots, \}$ be a node ID of the branch-and-bound tree 
where $0$ denotes the root node, 
$2 b^\prime + 1$ the left branch node of tree node $b^\prime$ 
and $2 b^\prime + 2$ the right branch node. 
An unvisited node set of the tree during the branch-and-bound process 
is denoted by $B$. 
At tree node $b$, we define $\overline{W_b}$ as 
$\overline{W_b} \subseteq \{ \{ i, i^\prime \} \mid i, i^\prime \in V \}$, 
i.e., a subset of all unordered pairs of the graph vertices. 
When $\{ i_1, i_2 \} \in \overline{W_b}$, 
we impose the left branching rule 
that $i_1$ and $i_2$ 
must belong to an identical possible community. 
Similarly, we define $\underline{W_b}$ 
and impose the right branching rule 
for $\{ i_1, i_2 \} \in \underline{W_b}$ 
that $i_1$ and $i_2$ 
must belong to a different possible community.

\subsection{Set-packing relaxation of restricted master}
\label{sec:BranchAndPrice_RestrictedMaster}
\mbox{ }

Straightforward column generation applied to the set-partitioning problem 
unfortunately requires much computation time 
for large instances due to degeneracy (in the LP context), 
as \citet{Lubbecke.2005} pointed out. 
For a set-partitioning-based minimization problem 
in the field of scheduling, 
\citet{Sato.2012} gave the set-covering relaxation 
to overcome the disadvantage. 
This technique first replaces the set-partitioning constraint 
with the set-covering constraint for all elements of the set. 
When the column generation converges, 
the technique focuses on the solution to the set-covering-relaxed LP 
and the set-covering-relaxed constraint set. 
For each element of the set, 
the constraint is reset 
if the value of its left-hand side exceeds that of its right-hand side, 
and then the column generation process is resumed. 
It is repeated until the column generation 
for a combination of the set-partitioning and the set-covering constraints 
converges and all the elements are exactly covered. 
Although this approach is much simpler 
than stabilized column generation proposed by \citet{Merle.1999}, 
it contributed to enough computation time reduction 
for their scheduling problem instances. 

In this study, we apply the set-packing relaxation 
to the set-partitioning problem~\mpprobref{mpprob:P} 
in our branch-and-price framework 
(since we discuss a maximization problem). 
We expect that the contribution of the majority of communities 
detected as an optimal solution will take a positive value, 
and therefore the set-packing constraint will suffice 
for a large part of the vertices. 
Let $(b,\ell)$ be the $\ell$-th iteration at branch-and-bound tree node $b$. 
We also let $\mathcal{C}_{(b,\ell)}$ and $V^{=}_{(b,\ell)}$ 
be subsets of $\mathcal{C}$ and $V$, respectively. 
Then we define the set-packing relaxation~\mpprobref{mpprob:RP} 
as follows: 
\begin{mpproblem}{RP$^\leq_{(b,\ell)}$}
\label{mpprob:RP}
\begin{alignat*}{3}
 &\text{max.} & \quad \sum_{C \in \mathcal{C}_{(b,\ell)}} f_C u_C      &
     & \quad &                                           \\ 
 &\text{s.t.} & \quad \sum_{C \in \mathcal{C}_{(b,\ell)}} a_{iC} u_{C} &= 1 
     & \quad &\forall i \in V^{=}_{(b,\ell)}             \\ 
 &            & \quad \sum_{C \in \mathcal{C}_{(b,\ell)}} a_{iC} u_{C} &\leq 1 
     & \quad &\forall i \in V \setminus V^{=}_{(b,\ell)} \\ 
 &            & \quad u_C                                              &\geq 0 
     & \quad &\forall C \in \mathcal{C}_{(b,\ell)}.         
\end{alignat*}
\end{mpproblem}
We should note here, for every $(b,\ell)$, 
that $\mathcal{C}_{(b,\ell)}$ must not contain any column 
which does not satisfy the left and right branching rule 
given by $\overline{W_b}$ and $\underline{W_b}$. 

Let us recall here the original set-partitioning problem~\mpprobref{mpprob:P} 
and consider the following problem~\mpprobref{mpprob:Pbl} 
that substitutes $\mathcal{C}_{(b,\ell)}$ for $\mathcal{C}$: 
\begin{mpproblem}{P$_{(b,\ell)}$}
\label{mpprob:Pbl}
\begin{alignat*}{3}
 &\text{max.} & \quad \sum_{C \in \mathcal{C}_{(b,\ell)}} f_C u_C      &
     & \quad &                                      \\
 &\text{s.t.} & \quad \sum_{C \in \mathcal{C}_{(b,\ell)}} a_{iC} u_{C} &= 1 
     & \quad &\forall i \in V                       \\
 &            & \quad u_C                                              &\in \{0 ,1\} 
     & \quad &\forall C \in \mathcal{C}_{(b,\ell)}. 
\end{alignat*}
\end{mpproblem}
If the problem is feasible, 
which is expected to be true for $\mathcal{C}_{(b,\ell)}$ 
consisting of a sufficient amount and variations of columns, 
its optimal value is clearly a lower bound of \mpprobref{mpprob:P}.

\subsection{MILP subproblem and multiple cutting planes as columns}
\label{sec:BranchAndPrice_Subproblem}
\mbox{ }

Let \mpprobref{mpprob:RD} be the dual of~\mpprobref{mpprob:RP} 
and $\lambda_i$ for vertex $i \in V$ be the dual variable. 
Then the problem is written as 
\begin{mpproblem}{RD$^\leq_{(b,\ell)}$}
\label{mpprob:RD}
\begin{alignat*}{3}
 &\text{min.} & \quad \sum_{i \in V} \lambda_i        &
     & \quad &                                            \\
 &\text{s.t.} & \quad \sum_{i \in V} a_{iC} \lambda_i &\geq f_C 
     & \quad &\forall C \in \mathcal{C}_{(b,\ell)}        \\
 &            & \quad \lambda_i                       &\in \Re 
     & \quad &\forall i \in V^{=}_{(b,\ell)}              \\
 &            & \quad \lambda_i                       &\geq 0 
     & \quad &\forall i \in V \setminus V^{=}_{(b,\ell)}. 
\end{alignat*}
\end{mpproblem}
We can see that the set-packing relaxation restricts the dual variables in sign, 
and removes the restriction for a subset of $V$ at some $(b,\ell)$. 
Such techniques are also reviewed in \citet{Lubbecke.2005}. 

We define $\lambda^\ast_{i,(b,\ell)}$ for $i \in V$ 
as the dual price of the constraint for $i$ 
at an optimal solution to \mpprobref{mpprob:RP} 
or as an optimal solution to \mpprobref{mpprob:RD}. 
Then the column generation subproblem~\mpprobref{mpprob:S} is given by 
\begin{mpproblem}{S$_{(b,\ell)}$}
\label{mpprob:S}
\begin{alignat*}{2}
 &\text{find}      & \quad &\mbox{\boldmath $x$} \in \{0, 1\}^V \setminus \{ 0^V \} \\
 &\text{such that} & \quad &\frac{  4 \sum_{\{i,j\} \in E} x_{i} x_{j} 
                                  - \sum_{i \in V} d_i x_{i}}
                                 {\sum_{i \in V} x_{i}} 
                            - \sum_{i \in V} \lambda^\ast_{i,(b,\ell)} x_{i} > 0    \\
 &                 & \quad &x_{i_1} - x_{i_2} = 0 
                                \quad \forall \{ i_1, i_2 \} \in \overline{W_b}     \\
 &                 & \quad &x_{i_1} + x_{i_2} \leq 1 
                                \quad \forall \{ i_1, i_2 \} \in \underline{W_b}.   
\end{alignat*}
\end{mpproblem}
Note that the last two constraints correspond to the branching rule. 
The former constraint indicates, for a pair of vertices in $\overline{W_b}$, 
that any column to be generated is not allowed 
to contain exactly either one of them 
since they must belong to an identical possible community. 
The latter constraint shows, for a pair in $\underline{W_b}$, 
that any column to be generated is not allowed 
to contain the both at the same time 
since they must belong to a different possible community. 

Adding not merely one column, or one cutting plane, 
but multiple ones at $(b,\ell)$ which may complement well each other 
will more likely contribute to fast convergence 
of the whole column generation process. 
\citet{Izunaga.2017} introduced the multiple-cutting-planes-at-a-time technique 
for its column generation subproblem of the modularity maximization. 
This technique first solves the original subproblem and obtain a column. 
It then removes the vertices included in the column 
from the whole vertex set of the subproblem, 
and solves the subproblem again. 
This procedure is repeated 
until the subproblem does not provide any column 
which may improve the objective function value 
of the corresponding restricted master problem 
or all the vertices are removed. 
This simple approach dramatically improved computation time 
of the modularity maximization 
for large instances solved in their paper. 

We solve \mpprobref{mpprob:S} 
by formulating it as a MILP problem, 
and apply the multiple-cutting-planes-at-a-time technique to the formulation. 
Let $q \in \{1, 2, \ldots, \}$ be an iteration counter of finding a column 
and $(b, \ell, q)$ be the $q$-th iteration at $(b,\ell)$. 
We give the column generation optimization subproblem~\mpprobref{mpprob:SS} 
below: 
\begin{mpproblem}{S$^{\textsf{MILP}}_{(b,\ell,q)}$}
\label{mpprob:SS}
\begin{alignat}{3}
 &\text{max.}        & \quad   4 \sum_{ \{ i,j \} \in E} w_{ij} 
                             - \sum_{i \in V} d_i y_{i} 
                             - \sum_{i \in V} \lambda^\ast_{i,(b,\ell)} x_{i} 
                                                  &               
     & \quad &                                           \label{eqn:SS_Obj}
\end{alignat}
\begin{alignat}{3}
 &\text{s.t.} \;\;\; & \quad \sum_{i \in V} y_{i} &=1             
     & \quad &                                           \label{eqn:SS_Con01} \\
 &                   & \quad 0 \leq s - y_{i}     &\leq 1 - x_{i} 
     & \quad &\forall i \in V                            \label{eqn:SS_Con02} \\
 &                   & \quad y_{i}                &\leq x_{i}     
     & \quad &\forall i \in V                            \label{eqn:SS_Con03} \\
 &                   & \quad w_{ij}               &\leq y_{i}     
     & \quad &\forall \{ i, j\} \in E                    \label{eqn:SS_Con04} \\
 &                   & \quad w_{ij}               &\leq y_{j}     
     & \quad &\forall \{ i, j\} \in E                    \label{eqn:SS_Con05} \\
 &                   & \quad x_{i_1} - x_{i_2}    &= 0            
     & \quad &\forall \{ i_1, i_2 \} \in \overline{W_b}  \label{eqn:SS_Con06} \\
 &                   & \quad x_{i_1} + x_{i_2}    &\leq 1         
     & \quad &\forall \{ i_1, i_2 \} \in \underline{W_b} \label{eqn:SS_Con07} \\
 &                   & \quad x_{i}                &\in \{ 0, 1 \} 
     & \quad &\forall i \in V \setminus V^0_{(b,\ell,q)} \label{eqn:SS_Con08} \\
 &                   & \quad x_{i}                &= 0            
     & \quad &\forall i \in V^0_{(b,\ell,q)}             \label{eqn:SS_Con09} \\
 &                   & \quad y_{i}                &\geq 0         
     & \quad &\forall i \in V                            \label{eqn:SS_Con10} \\
 &                   & \quad w_{ij}               &\in \Re        
     & \quad &\forall \{ i, j\} \in E                    \label{eqn:SS_Con11} \\
 &                   & \quad s                    &\in \Re.       
     & \quad &                                           \label{eqn:SS_Con12} 
\end{alignat}
\end{mpproblem}
To discuss the relationship 
between \mpprobref{mpprob:S} and \mpprobref{mpprob:SS}, 
let us consider the case with $q = 1$, 
which means $V^0_{(b,\ell,q)} = \emptyset$. 
The constraints~\eqref{eqn:SS_Con02}, \eqref{eqn:SS_Con03}, 
\eqref{eqn:SS_Con08}, \eqref{eqn:SS_Con10} and \eqref{eqn:SS_Con12} 
imply that $y_i = s x_i$ holds; 
if $x_i = 1$ then $s = y_i$ and otherwise $0 \leq y_i \leq x_i = 0$. 
This fact along with the constraint~\eqref{eqn:SS_Con01} 
implies $s = 1 / \sum_{i \in V} x_i$, 
and therefore $y_i = x_i / \sum_{i \in V} x_i$ holds. 
Note that $x_i = 0$ for all $i \in V$ is not a feasible solution. 
From the constraints~\eqref{eqn:SS_Con04}, \eqref{eqn:SS_Con05} 
and \eqref{eqn:SS_Con11} as well as the objective function to be maximized, 
$w_{ij} = \min \{ y_i, y_j \}$ holds 
at an optimal solution to the problem or to its linear relaxation problem, 
which is equivalent to $w_{ij} = x_i x_j / \sum_{i \in V} x_i$. 
Hence we can say 
that solving \mpprobref{mpprob:SS} answers \mpprobref{mpprob:S} for $q = 1$, 
and we can find another solution, if it exists, for $q \geq 2$ 
(which means $V^0_{(b,\ell,q)} \not= \emptyset$). 
If two or more columns are generated by this approach, 
they have no common vertex. 
Hence, we expect that these columns will coexist 
in a good feasible solution of the original or the LP relaxation of 
the set-partitioning/set-packing problem.

\subsection{Overall procedure}
\label{sec:BranchAndPrice_Procedure}
\mbox{ }

Our branch-and-price approach is displayed in Procedure~\ref{alg:BP}. 
Subroutine~\ref{alg:SPR} is called from the main procedure, 
and Subroutine~\ref{alg:MCP} is done from the preceding subroutine. 
\floatname{algorithm}{Procedure}
\begin{algorithm}[tb]
\caption{\textsc{Branch-and-price-to-density-maximization($G$)}}
\label{alg:BP}
\begin{algorithmic}[1]
\State{$B := \{ 0\}, (\overline{W_0},\underline{W_0}) := (\emptyset, \emptyset), 
    V^{=} := \emptyset, (\Pi, \texttt{LB}) := (\emptyset, -\infty)$}
    \label{op:BP_01}
\State{$\mathcal{C}_{(0,1)} := \mbox{initial set of columns}$}
    \label{op:BP_02}
\While{$B \not= \emptyset$}
    \State{$b := \mbox{$B$.\textsf{removeone()}}$ 
        according to any branch-and-bound node selection rule}
        \label{op:BP_04}
    \State{$(\mathcal{C}_{(b,\ell)}, 
        \mbox{\boldmath $u^{\texttt{U}_b\ast}$}, \texttt{UB}_{b}, 
        \mbox{\boldmath $u^{\texttt{L}_b\ast}$}, \texttt{LB}_{b}, V^{=}) :=$ 
        \Call{Set-packing-relaxation}
            {$G, b, \overline{W_b}, \underline{W_b}, \mathcal{C}_{(b,1)}, V^{=}$}}
            \label{op:BP_05}
    \If{$\texttt{UB}_{b} < \texttt{LB}$} \Comment{(bounding)}
        \State{continue}
    \EndIf
    \If{$\texttt{LB}_{b} > \texttt{LB}$} \Comment{(new incumbent solution)}
        \State{$(\Pi, \texttt{LB}) 
            := (\{ C \in \mathcal{C}_{(b,\ell)} \mid u^{\texttt{L}_b\ast}_{C} = 1 \}, \texttt{LB}_{b})$}
    \EndIf
    \If{$\texttt{UB}_{b} > \texttt{LB}_{b}$} \Comment{(branching)}
        \State{$\{i_1, i_2\} := $ \mbox{any $\{i,i^\prime\} \in V^2$ such that} 
            $a_{i C}, a_{i C^\prime}, a_{i^\prime C^\prime} = 1$, 
            $a_{i^\prime C} = 0$, 
            $0 < u^{\texttt{U}_b\ast}_{C}, u^{\texttt{U}_b\ast}_{C^\prime} < 1$ 
            \mbox{for some} $C, C^\prime \in \mathcal{C}_{(b,\ell)}$}
            \label{op:BP_11}
        \State{$(\overline{W_{2b+1}}, \underline{W_{2b+1}}) 
            := (\overline{W_b} \cup \{ i_1, i_2 \}, \underline{W_b})$}
        \State{$\mathcal{C}_{(2b+1,1)} := 
            \{ C \in \mathcal{C}_{(b,\ell)} 
            \mid (i_1 \in C \wedge i_2 \in C) \vee (i_1 \not\in C \wedge i_2 \not\in C)) \}$}
            \label{op:BP_13}
        \State{$(\overline{W_{2b+2}},\underline{W_{2b+2}}) 
            := (\overline{W_b}, \underline{W_b} \cup \{ i_1, i_2 \})$}
        \State{$\mathcal{C}_{(2b+2,1)} := 
            \{ C \in \mathcal{C}_{(b,\ell)} 
            \mid i_1 \not\in C \vee i_2 \not\in C \}$}
            \label{op:BP_15}
        \State{\mbox{\textsf{$B$.add($2b+2, 2b+1$)}}}
    \EndIf
\EndWhile
\State{\Return{$\Pi$}}
\end{algorithmic}
\end{algorithm}
Operations~\ref{op:BP_01} and \ref{op:BP_02} of Procedure~\ref{alg:BP} 
are initialization. 
Let $V^{=}$ be a subset of $V$ 
and also let $i \in V^{=}$ denote 
that the constraint for $i$ is the set-partitioning one 
at the beginning of the column generation for each $b$. 
If we substitute $V^{=} := V$ for $V^{=} := \emptyset$, 
then the restricted master problem 
has the standard set-partitioning constraints only. 
Two symbols $\Pi$ and \texttt{LB} indicate 
an incumbent solution and its objective value, respectively. 
In our numerical experiments carried out in Section~\ref{sec:Results}, 
we take $\mathcal{C}_{(0,1)} := \{ \{ i \} \mid i \in V \}$ 
as the initial set of columns. 
Also, we pick branch-and-bound tree node $b$ 
according to a depth-first rule at Operation~\ref{op:BP_04}. 
The left node is chosen before the right node is done. 
At Operation~\ref{op:BP_05}, Subroutine~\ref{alg:SPR} returns 
fully generated column set $\mathcal{C}_{(b,\ell)}$ 
as well as upper and lower bound information on \mpprobref{mpprob:Pbl}. 
An optimal solution vector to the linear relaxation problem of \mpprobref{mpprob:Pbl} 
and its objective value 
are denoted by \mbox{\boldmath $u^{\texttt{U}_b\ast}$} and $\texttt{UB}_{b}$. 
Similarly, a lower bound solution vector to \mpprobref{mpprob:Pbl} 
and its objective value 
are given by \mbox{\boldmath $u^{\texttt{L}_b\ast}$} and $\texttt{LB}_{b}$. 
The set $V^{=}$ is also updated in the subroutine. 
The vector \mbox{\boldmath $u^{\texttt{L}_b\ast}$} is integral, 
whereas \mbox{\boldmath $u^{\texttt{U}_b\ast}$} is fractional 
if $\texttt{UB}_{b} > \texttt{LB}_{b}$ holds, i.e., 
there is a gap between the upper and lower bound. 
Operation~\ref{op:BP_11} starts the branching scheme 
described in Subsection~\ref{sec:BranchAndPrice_BranchAndPrice}. 
In our experiments, 
we simply search the relevant lists from their heads. 
We first do the vector list \mbox{\boldmath $u^{\texttt{U}_b\ast}$} 
from its head, and add the corresponding column to a temporal list 
for each variable whose value is fractional. 
Next, for the double loop of the temporal list and for the vertex set list, 
we check if the currently selected vertex is contained 
in both of the currently selected column pair 
or in only one of them. 
At Operations~\ref{op:BP_13} and \ref{op:BP_15}, we prepare 
initial column sets $\mathcal{C}_{(2b+1,1)}$ and $\mathcal{C}_{(2b+2,1)}$ 
for nodes $2b+1$ and $2b+2$, respectively. 
Only the columns satisfying 
the ``identical restrictions on subsets'' branching rule are selected. 
When the whole procedure terminates, $\Pi$ is output as an optimal solution.

\floatname{algorithm}{Subroutine}
\setcounter{algorithm}{0}
\begin{algorithm}[tb]
\caption{\textsc{Set-packing-relaxation%
    ($G, b, \overline{W_b}, \underline{W_b}, \mathcal{C}_{(b,1)}, V^{=}$)}}
\label{alg:SPR}
\begin{algorithmic}[1]
\State{$\ell := 1, V^{=}_{(b,1)} := V^{=}$}
    \label{op:SPR_01}
\Loop
    \State{Solve \mpprobref{mpprob:RP}}
        \label{op:SPR_03}
    \State{$(\mbox{\boldmath $u^{\ast}$}, \mbox{\boldmath $\lambda^\ast_{(b,\ell)}$}) 
        := \mbox{primal and dual optimal solution to \mpprobref{mpprob:RP}}$}
        \label{op:SPR_04}
    \State{$\widehat{\mathcal{C}} :=$ 
        \Call{Multiple-cutting-planes}{$G, b, \overline{W_b}, \underline{W_b}, \ell, 
            \mbox{\boldmath $\lambda^\ast_{(b,\ell)}$}$}}
        \label{op:SPR_05}
    \If{$\widehat{\mathcal{C}} \not= \emptyset$}
        \State{$\mathcal{C}_{(b,\ell+1)} 
            := \mathcal{C}_{(b,\ell)} \cup \widehat{\mathcal{C}}$}
        \State{$V^{=}_{(b,\ell+1)} := V^{=}_{(b,\ell)}$}
        \State{$\ell := \ell + 1$}
    \Else
        \State{$\widehat{V} := \{ i \in V \setminus V^{=}_{(b,\ell)} 
            \mid \sum_{C \in \mathcal{C}_{(b,\ell)}} a_{iC} u^{\ast}_{C} < 1 \}$}
            \label{op:SPR_11}
        \If{$\widehat{V} \not= \emptyset$}
            \State{$V^{=}_{(b,\ell+1)} := V^{=}_{(b,\ell)} \cup \widehat{V}$}
            \State{$\ell := \ell + 1$}
        \Else
            \State{$V^{=} := V^{=}_{(b,\ell)}$}
            \State{$(\mbox{\boldmath $u^{\texttt{U}_b\ast}$}, \texttt{UB}_{b}) 
                := \mbox{objective solution and value of \mpprobref{mpprob:RP}}$}
                \label{op:SPR_16}
            \If{$u^{\texttt{U}_b\ast}_C \in \{ 0, 1\} \;\; \forall C \in \mathcal{C}_{(b,\ell)}$}
                \State{$(\mbox{\boldmath $u^{\texttt{L}_b\ast}$}, \texttt{LB}_{b}) 
                    := (\mbox{\boldmath $u^{\texttt{U}_b\ast}$}, \texttt{UB}_{b})$}
            \Else
                \State{\mbox{Solve \mpprobref{mpprob:Pbl}}}
                    \label{op:SPR_20}
                \State{$(\mbox{\boldmath $u^{\texttt{L}_b\ast}$}, \texttt{LB}_{b}) 
                    := \mbox{objective solution and value of \mpprobref{mpprob:Pbl}}$}
                    \label{op:SPR_21}
            \EndIf
            \State{\Return{$(\mathcal{C}_{(b,\ell)}, 
                \mbox{\boldmath $u^{\texttt{U}_b\ast}$}, \texttt{UB}_{b}, 
                \mbox{\boldmath $u^{\texttt{L}_b\ast}$}, \texttt{LB}_{b}, V^{=})$}}
        \EndIf
    \EndIf
\EndLoop
\end{algorithmic}
\end{algorithm}
Subroutine~\ref{alg:SPR} corresponds to the content 
of Subsection~\ref{sec:BranchAndPrice_RestrictedMaster}. 
At Operation~\ref{op:SPR_01}, 
we let the set $V^{=}_{(b,1)}$ 
which appears in the restricted master problem~\mpprobref{mpprob:RP} 
be $V^{=}$. 
After solving the restricted master problem at Operation~\ref{op:SPR_03}, 
we let \mbox{\boldmath $u^{\ast}$} and \mbox{\boldmath $\lambda^\ast_{(b,\ell)}$} 
be its optimal primal and dual solution vectors at Operation~\ref{op:SPR_04}. 
We solve \mpprobref{mpprob:RP} 
by an optimization solver in our numerical experiments. 
Note that an interior point method is applied to this problem 
according to an indication by \citet{Vanderbeck.2005} 
that fewer iterations are required 
for column generation to terminate 
if an analytic center of the optimal face is provided as the solution. 
At Operation~\ref{op:SPR_05}, columns are generated by Subroutine~\ref{alg:MCP} 
and the generated column set is denoted by $\widehat{\mathcal{C}}$. 
If any column is generated, 
then we update the column set of \mpprobref{mpprob:RP} and solve it again. 
Otherwise, we define set $\widehat{V}$ and 
focus on each of the set-packing constraint set 
as well as the solution value \mbox{\boldmath $u^{\ast}$} 
at Operation~\ref{op:SPR_11}. 
If the value of its left-hand side is less than that of its right-hand side, 
the vertex index which corresponds to such constraint 
is collected in $\widehat{V}$. 
We make the set-packing constraint for each of the elements of $\widehat{V}$ 
the equality one, and go to Operation~\ref{op:SPR_03}. 
If $\widehat{V}$ is empty, 
then it shows that all the vertices are exactly covered. 
Recall here that $\widehat{\mathcal{C}}$ is also empty at this iteration, 
i.e., there is no unknown column which may contribute 
to the improvement of the objective value of \mpprobref{mpprob:RP}. 
The two facts indicate the convergence of the column generation 
at the branch-and-bound tree node $b$, 
and the upper bound of \mpprobref{mpprob:Pbl} is obtained 
at Operation~\ref{op:SPR_16}. 
If \mbox{\boldmath $u^{\texttt{U}_b\ast}$} is an integral solution, 
then we have fortunately found the optimal solution to \mpprobref{mpprob:Pbl} 
at $b$. There is no gap between the upper and lower bound. 
Otherwise, we solve \mpprobref{mpprob:Pbl} 
to find an integral solution at Operation~\ref{op:SPR_20}. 
The problem is also solved to optimality by the optimization solver. 
Finally, the column set at the termination of the column generation, 
the upper bound solution as well as its objective value 
and the lower bound solution as well as its objective value 
at the branch-and-bound tree node are output. 

\begin{algorithm}[tb]
\caption{\textsc{Multiple-cutting-planes(%
    $G, b, \overline{W_b}, \underline{W_b}, \ell, 
    \mbox{\boldmath $\lambda^\ast_{(b,\ell)}$)}$}}
\label{alg:MCP}
\begin{algorithmic}[1]
\State{$q := 1, V^0_{(b,\ell,1)} := \emptyset, 
    \widehat{\mathcal{C}} := \emptyset$}
    \label{op:MCP_01}
\Loop
    \State{\mbox{Find any solution \mbox{\boldmath $\widehat{x}$} to \mpprobref{mpprob:SS} with positive objective value}}
        \label{op:MCP_03}
    \If{\mbox{\boldmath $\widehat{x}$} is found}
        \State{$\widehat{\mathcal{C}} := \widehat{\mathcal{C}} \cup \{ \{ i \in V \setminus V^0_{(b,\ell,q)} \mid \widehat{x}_i = 1 \} \}$}
        \State{$V^0_{(b,\ell,q+1)} := V^0_{(b,\ell,q)} \cup \{ i \in V \setminus V^0_{(b,\ell,q)} \mid \widehat{x}_i = 1 \}$}
        \State{$q := q + 1$}
    \Else
        \State{\Return{$\widehat{\mathcal{C}}$}}
    \EndIf
\EndLoop
\end{algorithmic}
\end{algorithm}
The content of Subsection~\ref{sec:BranchAndPrice_Subproblem} 
is coded in Subroutine~\ref{alg:MCP}. 
At Operation~\ref{op:MCP_01}, 
let $\widehat{\mathcal{C}}$ be a set to which we add new columns. 
At Operation~\ref{op:MCP_03}, 
the column generation MILP subproblem~\mpprobref{mpprob:SS} 
is solved by the optimization solver 
in our numerical experiments. 
Note that any solution to \mpprobref{mpprob:SS} 
with a positive objective value, 
denoted by \mbox{\boldmath $\widehat{x}$}, 
suffices as a new column to be added to \mpprobref{mpprob:RP}. 
We add, in our experiments, 
the first incumbent solution with a positive objective value 
found in the branch-and-bound process of the MILP. 
Here it is fair not to rely heavily on heuristics 
implemented in the solver to find a feasible mixed integer solution, 
hence we tune its parameters 
and expect an LP optimal solution 
satisfying the integral constraint 
at a branch-and-bound tree node of the MILP. 
This approach requires less computation time 
than searching for an optimal solution. 
On the other hand, it may increase 
the total number of the column generation iterations $\ell$. 
This discussion will be meaningless 
if there exists no solution with a positive objective value; 
in such case we have to optimize \mpprobref{mpprob:SS} 
to prove the nonexistence. 
After solving \mpprobref{mpprob:SS} 
we remove all the element of \mbox{\boldmath $\widehat{x}$} 
from the formulation, and solve the problem again. 
We get out of the loop 
if there exists no solution with a positive objective value 
or all the vertices in $V$ are removed. 
The set of columns to be added at the next column generation iteration 
is returned as the output. 

\section{Numerical Results}
\label{sec:Results}
\subsection{Instances and computational environment}
\mbox{ }

We solve several real graph instances seen in the literature 
by our exact branch-and-price approach. 
For comparison, 
we also solve them 
by the best MILP formulation called MDB by \citet{Costa.2015}, 
and by our branch-and-price algorithm 
in which the column generation subproblem is 
\mpprobref{mpprob:SS-IQP} modeled by \citet{Santiago.InPress} 
with the branching constraints \eqref{eqn:SS_Con06} and \eqref{eqn:SS_Con07}. 
We calculate the upper bound value of the contribution of a community 
required as input of MDB 
by the parametric algorithm by \citet{Dinkelbach.1967}, 
as \citet{Izunaga.2016} did. 

Table~\ref{tab:Instances} summarizes the instances. 
They are from 
\citet{Costa.2015} (IDs~01--10), 
\citet{Costa.2016} (IDs~11, 13--15), 
\citet{Santiago.InPress} (IDs~12, 16) and 
\citet{Santiago.2017} (IDs~17, 18), respectively. 
For each proven optimal (IDs~01--10, 12, 16) 
or best-known heuristic (IDs~11, 13--15, 17, 18) solution, 
its objective value and the number of detected communities 
are indicated as ``Best-known $D(\Pi)$'' and ``Best-known $|\Pi|$.''
%
%
\begin{table}[tb]
\centering
\caption{Instances.}
\label{tab:Instances}
\begin{tabular}{clrrr@{.}l@{, }r} \hline
ID & \multicolumn{1}{c}{Name} & \multicolumn{1}{c}{$n$} & \multicolumn{1}{c}{$m$} 
    & \multicolumn{3}{c}{Best-known} \\ 
 & & & & \multicolumn{2}{c@{, }}{$D(\Pi)$} & $|\Pi|$~{} \\ \hline\hline 
01 & Strike             &  24 &  38 &  8&86111 &  4$^\bullet$ \\
02 & Galesburg F        &  31 &  63 &  8&28571 &  3$^\bullet$ \\
03 & Galesburg D        &  31 &  67 &  6&92692 &  3$^\bullet$ \\
04 & Karate             &  34 &  78 &  7&8451  &  3$^\bullet$ \\
05 & Korea 1            &  35 &  69 & 10&9667  &  5$^\bullet$ \\
06 & Korea 2            &  35 &  84 & 11&143   &  5$^\bullet$ \\
07 & Mexico             &  35 & 117 &  8&71806 &  3$^\bullet$ \\
08 & Sawmill            &  36 &  62 &  8&62338 &  4$^\bullet$ \\
09 & Dolphins small     &  40 &  70 & 13&0519  &  8$^\bullet$ \\
10 & Journal index      &  40 & 189 & 17&8     &  4$^\bullet$ \\
11 & Graph              &  60 & 114 &  9&57875 &  7~{}        \\
12 & Dolphins           &  62 & 159 & 12&1252  &  5$^\bullet$ \\
13 & Les Mis\'{e}rables &  77 & 254 & 24&5339  &  9~{}        \\
14 & A00 main           &  83 & 125 & 13&3731  & 11~{}        \\
15 & Protein p53        & 104 & 226 & 12&9895  &  8~{}        \\
16 & Political books    & 105 & 441 & 21&9652  &  7$^\bullet$ \\
17 & Adjnoun            & 112 & 425 &  7&651   &  2~{}        \\
18 & Football           & 115 & 613 & 44&340   & 10~{}        \\ \hline
\multicolumn{7}{r}{$^\bullet$: proven optimal solution.} 
\end{tabular}
\end{table}

The programs are implemented in Python~3.5.2, calling the Python API 
of Gurobi Optimizer~7.0.2 (developed by \citet{Gurobi.2017}) 
to solve the LP, ILP, MILP and IQP problems. 
The instances are solved on a 64-bit Windows~10 PC 
having a Core~i7-6700 CPU (fore cores, eight threads, 3.4--4.0~GHz) 
and 32~GB RAM (the actual usage is less than 3~GB). 
We stop the algorithms after 3,600 seconds (one hour) 
if the corresponding instance is not solved within the time limit. 
In a case where our Procedure~\ref{alg:BP} reaches the limit, 
we collect all columns generated until then 
and solve \mpprobref{mpprob:Pbl}, 
giving a lower bound solution.

\subsection{Solved instances}
\mbox{ }

\begin{figure}[tb]
\centering
\begin{tikzpicture}
    \tikzset{ block/.style={rectangle,text width=74, draw=black} };
    \node[very thick, circle,draw=black] {\large 0} [level distance=1.5cm, sibling distance=2cm, edge from parent/.style={->,draw}, very thick]
    child{ node[very thick,circle, draw=black]{\large 1} }
    child{ node[very thick,circle, draw=black]{\large 2} 
        child{ node[semithick,double,circle, draw=black]{\large 5} }
        child{ node[very thick,circle, draw=black]{\large 6} }
    };
    \node (branch1) at (0,-0.97) {\small $\{ 1,16 \}$};
    \node (branch2) at (1,-2.47) {\small $\{ 1,12 \}$};
    \node[block] (n0) at (-2.45,0) {\small $\texttt{UB}_{0}=13.22153$ $\texttt{LB}_{0}=13.21122$};
    \draw (-0.4,0)--(-1,0);
    \node[block] (n1) at (-3.45,-1.5) {\small $\texttt{UB}_{1}=13.19581$ $\texttt{LB}_{1}=13.19581$};
    \draw (-1.4,-1.5)--(-2,-1.5);
    \node[block] (n2) at (3.85,-1.5) {\small $\texttt{UB}_{2}=13.22153$ $\texttt{LB}_{2}=13.21122$};
    \draw (1.4,-1.5)--(2.4,-1.5);
    \node[block] (n5) at (-2.45,-3) {\small $\texttt{UB}_{5}=13.21433$ $\texttt{LB}_{5}=13.21433$};
    \draw (-0.4,-3)--(-1,-3);
    \node[block] (n6) at (4.45,-3) {\small $\texttt{UB}_{6}=13.19581$ $\texttt{LB}_{6}=13.19581$};
    \draw (2.4,-3)--(3,-3);
    \draw[dotted,very thick] (0,-0.7)--(1,0)--(4.2,0);
    \draw[dotted,very thick] (1.0,-2.2)--(3.0,-0.1)--(4.2,-0.1);
    \node (strategy) at (2.6,0.3) {\small Branching strategy};
\end{tikzpicture}
\caption{Branch-and-bound tree of Protein p53 instance (ID~15).}
\label{fig:BB_Pp}
\end{figure}
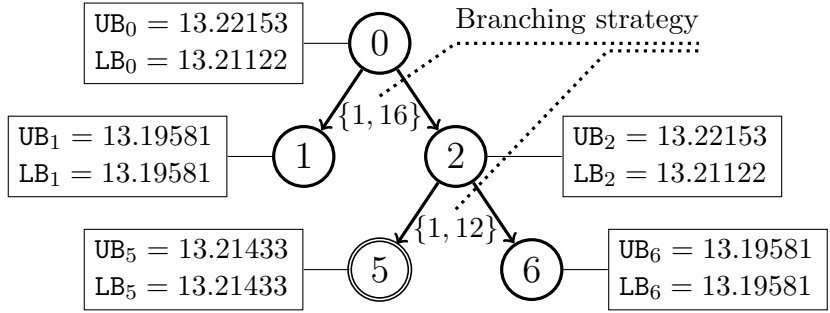

Table~\ref{tab:Solutions} 
shows an optimal modularity density value 
and the corresponding number of communities 
obtained by our Procedure~\ref{alg:BP} for each instance. 
We let `\#$b$' in the table 
be the number of branch-and-bound tree nodes processed. 
\begin{table}[tb]
\centering
\caption{Solutions by our approach.}
\label{tab:Solutions}
\begin{tabular}{c@{\hspace{4mm}}c@{\hspace{4mm}}r@{, }r} \hline
ID & \#$b$ 
   & \multicolumn{2}{c}{Optimal}       \\ 
   & & \multicolumn{1}{c@{, }}{$D(\Pi)$} & $|\Pi|$~{} \\ \hline\hline 
01 & 1 &  8.86111 &  4~{}              \\
02 & 1 &  8.28571 &  3~{}              \\
03 & 1 &  6.92692 &  3~{}              \\
04 & 1 &  7.84510 &  3~{}              \\
05 & 1 & 10.96667 &  5~{}              \\
06 & 1 & 11.14301 &  5~{}              \\
07 & 1 &  8.71806 &  3~{}              \\
08 & 1 &  8.62338 &  4~{}              \\
09 & 1 & 13.05195 &  8~{}              \\
10 & 1 & 17.80000 &  4~{}              \\
11 & 1 &  9.75238 &  7$^\ast$          \\
12 & 1 & 12.12523 &  5~{}              \\
13 & 1 & 24.54744 &  8$^\ast$          \\
14 & 1 & 13.48249 & 12$^\ast$          \\
15 & 5 & 13.21433 &  9$^\ast$          \\
16 & 1 & 21.96515 &  7~{}              \\
17 & \multicolumn{3}{c}{--$^\diamond$} \\
18 & \multicolumn{3}{c}{--$^\diamond$} \\ \hline
\multicolumn{4}{l}{$^\ast$: newly found solution.} \\
\multicolumn{4}{l}{$^\diamond$: timeout of 3,600 seconds.} 
\end{tabular}
\end{table}
We have found new and optimal solutions 
for Graph (ID~11), Les~Mis\'{e}rables (ID~13),
A00~main (ID~14) and Protein~p53 (ID~15) instances. 
Above all, the result for the last instance is remarkable; 
the column generation at the root node of the branch-and-bound tree 
has provided a fractional upper bound solution 
and five branch-and-bound tree nodes have been processed. 
Figure~\ref{fig:BB_Pp} shows the branch-and-bound tree. 
This result justifies the necessity of our branch-and-price approach. 
The instances having up to 105 vertices have been solved. 
Instances IDs~17 and 18, 
which consist of 112 and 115 vertices respectively, 
have been shown to be intractable 
after 3,600 seconds of the computation. 

\subsection{Computation time}
\mbox{ }

The computation time depending the solution methods 
is summarized in Table~\ref{tab:Time}. 
The symbol `MDB' means the best formulation by \citet{Costa.2015}, 
`BP-\mpprobref{mpprob:SS-IQP}' 
our branch-and-price approach combined with 
the column generation subproblem formulation 
by \citet{Santiago.InPress} 
and `BP-\mpprobref{mpprob:SS}' our approach 
with the MILP subproblem formulation. 
In the last approach, 
`SRP'/`No-SRP' and `MCP'/`No-MCP' indicate 
that the set-packing relaxation 
and the multiple-cutting-planes-at-a-time techniques 
are enabled/disabled, respectively. 
The best result for each instance is marked in bold. 
Note that the computation time of MDB 
includes that of the upper bound calculation by the parametric algorithm, 
which has been solved instantly for each of all the instances. 
\begin{table}[tb]
\centering
\caption{Computation time (in seconds).}
\label{tab:Time}
\begin{tabular}{crrrrrr} \hline
ID & \multicolumn{1}{c}{MDB} & \multicolumn{1}{c}{BP-\mpprobref{mpprob:SS-IQP}} 
& \multicolumn{4}{c}{BP-\mpprobref{mpprob:SS}} \\
   & & \multicolumn{1}{c}{SPR} & \multicolumn{1}{c}{SPR} & \multicolumn{1}{c}{SPR}    
& \multicolumn{1}{c}{No-SPR} & \multicolumn{1}{c}{No-SPR} \\
   & & \multicolumn{1}{c}{MCP} & \multicolumn{1}{c}{MCP} & \multicolumn{1}{c}{No-MCP} 
& \multicolumn{1}{c}{MCP}    & \multicolumn{1}{c}{No-MCP} \\ \hline\hline 
01 &           0.6 &           0.9 &           0.4 &           \textbf{0.3} &           1.0 &           1.5 \\
02 &           \textbf{0.5} &           3.7 &           0.7 &           0.6 &           3.4 &           5.0 \\
03 &           1.1 &           4.5 &           \textbf{1.0} &           \textbf{1.0} &           4.8 &           7.3 \\
04 &           \textbf{0.5} &           4.3 &           1.3 &           1.0 &           7.9 &           8.9 \\
05 &           8.8 &          11.3 &           1.0 &           \textbf{0.9} &           6.0 &          10.8 \\
06 &          88.1 &           3.8 &           1.3 &           \textbf{1.2} &           6.2 &          10.0 \\
07 &           9.6 &          12.5 &           \textbf{3.2} &           3.5 &          11.9 &          20.7 \\
08 &           3.4 &           7.0 &           \textbf{0.8} &           \textbf{0.8} &           3.8 &          12.2 \\
09 &        2848.4 &          15.8 &           0.8 &           \textbf{0.6} &           5.1 &          15.4 \\
10 &         651.0 &           6.4 &           4.7 &           \textbf{3.1} &          25.3 &          44.3 \\
11 & --$^\diamond$ &        1194.4 &           \textbf{5.4} &           7.8 &          46.5 &         190.4 \\
12 & --$^\diamond$ & --$^\diamond$ &          \textbf{18.4} &          18.6 &          84.2 &         419.3 \\
13 & --$^\diamond$ & --$^\diamond$ &          71.6 &          \textbf{60.7} &         154.6 &         335.0 \\
14 & --$^\diamond$ & --$^\diamond$ &           \textbf{3.8} &          15.6 &          82.5 &        1436.5 \\
15 & --$^\diamond$ & --$^\diamond$ &         223.5 &         \textbf{134.4} &        1227.5 & --$^\diamond$ \\
16 & --$^\diamond$ & --$^\diamond$ &         \textbf{242.8} &         498.2 & --$^\diamond$ & --$^\diamond$ \\
17 & --$^\diamond$ & --$^\diamond$ & --$^\diamond$ & --$^\diamond$ & --$^\diamond$ & --$^\diamond$ \\
18 & --$^\diamond$ & --$^\diamond$ & --$^\diamond$ & --$^\diamond$ & --$^\diamond$ & --$^\diamond$ \\ \hline
\multicolumn{7}{r}{$^\diamond$: timeout of 3,600 seconds.} 
\end{tabular}
\end{table}
As a whole, column generation to the modularity density maximization 
has outperformed MDB 
for the instances having 40 or more vertices (IDs~09--16), 
and the MILP formulation of the column generation subproblem 
has been easier to solve than the IQP formulation has been. 
Note here that our branch-and-price approach as well as MDB 
is deterministic 
and the computation time is not affected by any stochastic behavior. 
The set-packing relaxation 
has dramatically reduced the computation time 
for the instances having 60 or more vertices (IDs~11--14) 
and has enabled us 
to solve the instances having over 100 instances (IDs 15 and 16) 
within the time limit. 
The multiple-cutting-planes-at-a-time technique 
applied to the standard set-partitioning column generation process 
has been shown to be quite effective, 
as it was shown on the modularity maximization by \citet{Izunaga.2017}. 
On the other hand, 
the positive effect of this techniques 
when combined with the set-packing relaxation 
has depended on the instances. 
We should note, nevertheless, 
that the largest Political books instance (ID~16) 
among the successfully solved ones 
has been optimized in around four minutes 
by applying the both techniques. 

Table~\ref{tab:Iterations} shows 
details of our column generation results 
with the MILP subproblem formulation. 
In this table, 
we let `$\sum \ell$' be the total number of the column generation iterations 
over all the branch-and-bound tree nodes, 
`$\sum |\mathcal{C}_{(b,\ell)}|$' the total number of columns generated 
over all the nodes 
and `${V^{=}}$' the total number of vertices 
whose corresponding set-packing constraint 
has been changed to the set-partitioning constraint 
in the algorithm, respectively. 
The best result 
in terms of less numbers of $\sum \ell$ or $\sum |\mathcal{C}_{(b,\ell)}|$ 
for each instance is marked in bold. 
\begin{table}[tb]
\centering
\caption{Column generation iteration results.}
\label{tab:Iterations}
\begin{tabular}{cr@{, }r@{, }rr@{, }r@{, }rr@{, }rr@{, }r} \hline
ID & \multicolumn{10}{c}{CG-\mpprobref{mpprob:SS}} \\
   & \multicolumn{3}{c}{SPR}    & \multicolumn{3}{c}{SPR}    
   & \multicolumn{2}{c}{No-SPR} & \multicolumn{2}{c}{No-SPR} \\
   & \multicolumn{3}{c}{MCP}    & \multicolumn{3}{c}{No-MCP} 
   & \multicolumn{2}{c}{MCP}    & \multicolumn{2}{c}{No-MCP} \\ 
   & \multicolumn{6}{c}{($\sum \ell$, $\sum |\mathcal{C}_{(b,\ell)}|$, $|V^{=}|$)}
   & \multicolumn{4}{c}{($\sum \ell$, $\sum |\mathcal{C}_{(b,\ell)}|$)} \\ \hline\hline 
01 &  (\textbf{21} &  \textbf{52} &  1)~{}         &  (34 &  56 &  1)~{}         &  ( 64 &  104)~{}         &   (120 &   143)~{}         \\
02 &  (\textbf{25} &  72 &  1)~{}         &  (36 &  \textbf{65} &  1)~{}         &  (153 &  205)~{}         &   (331 &   361)~{}         \\
03 &  (\textbf{36} &  \textbf{76} &  1)~{}         &  (48 &  77 &  1)~{}         &  (202 &  260)~{}         &   (432 &   462)~{}         \\
04 &  (\textbf{40} &  \textbf{83} &  1)~{}         &  (55 &  87 &  1)~{}         &  (317 &  377)~{}         &   (481 &   514)~{}         \\
05 &  (\textbf{35} &  \textbf{84} &  3)~{}         &  (55 &  88 &  3)~{}         &  (217 &  295)~{}         &   (562 &   596)~{}         \\
06 &  (\textbf{31} &  84 &  1)~{}         &  (45 &  \textbf{78} &  1)~{}         &  (231 &  300)~{}         &   (499 &   533)~{}         \\
07 &  (\textbf{46} &  \textbf{89} &  2)~{}         &  (57 &  90 &  2)~{}         &  (325 &  384)~{}         &   (635 &   669)~{}         \\
08 &  (\textbf{25} &  87 &  0)~{}         &  (48 &  \textbf{83} &  0)~{}         &  (164 &  244)~{}         &   (688 &   723)~{}         \\
09 &  (\textbf{21} &  90 &  0)~{}         &  (47 &  \textbf{86} &  0)~{}         &  (189 &  285)~{}         &   (850 &   889)~{}         \\
10 &  (\textbf{35} &  84 &  1)~{}         &  (42 &  \textbf{80} &  1)~{}         &  (483 &  560)~{}         &  (1056 &  1095)~{}         \\
11 &  (\textbf{64} & \textbf{169} &  7)~{}         & (134 & 192 &  7)~{}         &  (663 &  878)~{}         &  (3222 &  3281)~{}         \\
12 &  (\textbf{77} & \textbf{188} &  5)~{}         & (147 & 207 &  5)~{}         &  (953 & 1107)~{}         &  (5140 &  5201)~{}         \\
13 & (\textbf{107} & \textbf{227} & 12)~{}         & (171 & 245 & 12)~{}         &  (809 &  960)~{}         &  (2925 &  3001)~{}         \\
14 & ( \textbf{47} & \textbf{189} &  1)~{}         & (135 & 216 &  1)~{}         &  (907 & 1145)~{}         & (10234 & 10316)~{}         \\
15 & (\textbf{143} & \textbf{329} &  1)~{}         & (231 & 331 &  1)~{}         & (2086 & 2554)~{}         & (14381 & 14484)$^\diamond$ \\
16 & (\textbf{117} & \textbf{311} &  6)~{}         & (327 & 430 &  6)~{}         & (4497 & 5372)$^\diamond$ & (10527 & 10631)$^\diamond$ \\
17 & (\textbf{112} & \textbf{223} &  0)$^\diamond$ & (\textbf{112} & \textbf{223} &  0)$^\diamond$ & (6284 & 6424)$^\diamond$ & ( 8372 &  8483)$^\diamond$ \\
18 & ( \textbf{60} & \textbf{191} &  0)$^\diamond$ & ( 98 & 212 &  0)$^\diamond$ & (4135 & 4392)$^\diamond$ & ( 7991 &  8105)$^\diamond$ \\ \hline
\multicolumn{11}{r}{$^\diamond$: timeout of 3,600 seconds.} 
\end{tabular}
\end{table}
The set-packing relaxation or the multiple-cutting-planes-at-a-time technique 
has dramatically reduced the number of column generation iterations and generated columns. 
When it comes to the total number of generated columns, 
the positive effect of the combination of the techniques has depended on the instances. 
A low percentage of the vertices have required the set-partitioning constraint, 
which supports the set-packing relaxation. 
Here let us focus on the unsolved instance ID~17 
even though we have applied the set-packing relaxation. 
With or without the multiple-cutting-planes-at-a-time technique, 
\mpprobref{mpprob:RP} has been solved instantly for every $(b,\ell)$ 
and \mpprobref{mpprob:SS} in short time until $(b,\ell)$ has reached $(0,111)$ 
(and for every $q$ when we have enabled the multiple-cutting-planes-at-a-time). 
The column generation subproblem for $(b,\ell) = (0,112)$ (and for $q=1$), 
however, has not been able to find even a solution 
with a positive objective value in almost one hour. 
This is caused by the value of $\lambda^\ast_{i,(b,\ell)}$, 
i.e., the dual solution of \mpprobref{mpprob:RP}, 
that is obtained by an interior point method. 
We have also tried a dual simplex method, 
which in turn has caused an extremely slow improvement 
of the objective value of \mpprobref{mpprob:RP} over $(b,\ell)$. 
The midpoint of the two dual solution values has not resolved the issue. 
A similar result has been observed for instance ID~18. 

\begin{figure}[tb]
\centering
\includegraphics[scale=0.78]{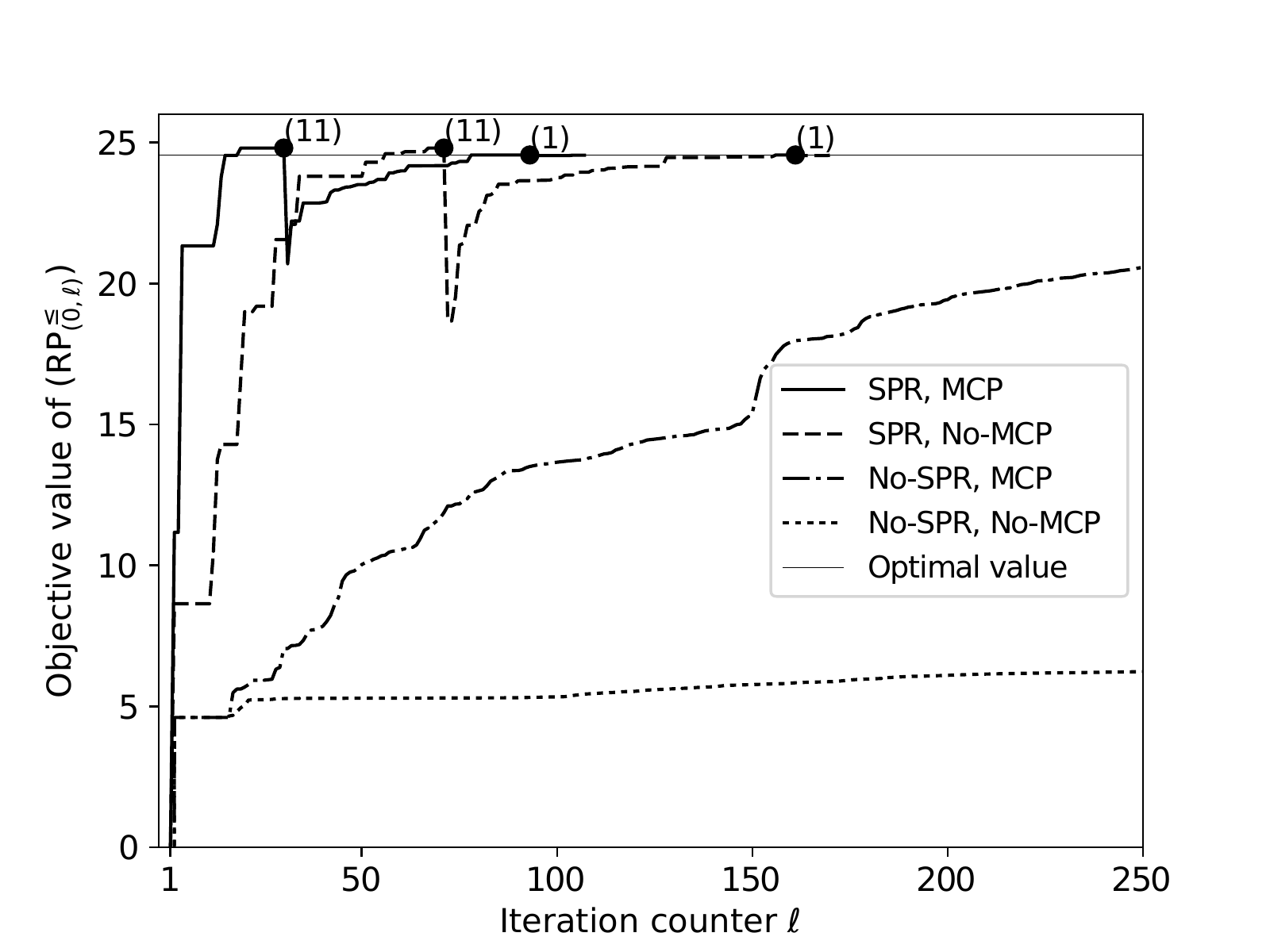}
\caption{Column generation for Les Mis\'{e}rables instance (ID~13).}
\label{fig:CG_LM}
\end{figure}

Figure~\ref{fig:CG_LM} plots, for Les Mis\'{e}rables instance (ID~13), 
the objective value of \mpprobref{mpprob:RP} 
for each iteration counter value $\ell$ 
($b = 0$ since the instance has been solved to optimality at the root node). 
The numbers in parentheses 
indicates $\widehat{V}$ in Subroutine~\ref{alg:SPR}, 
i.e., the number of vertices whose corresponding constraint 
has been changed from the set-packing to the set-partitioning one 
at $\ell+1$. 
This figure clearly indicates 
the positive effect of the set-packing relaxation. 
For this case, 
our algorithm with the set-packing relaxation 
and without the multiple-cutting-planes-at-a-time 
has required the least computation time 
in spite of more iterations than that with the both techniques. 
That is since \mpprobref{mpprob:SS} is solved only once for each $\ell$, 
whereas the multiple-cutting-planes-at-a-time technique 
has to show 
that no column to be added is left for some $q^\prime$ 
after removing the vertices found at $q = 1 \ldots, q^\prime - 1$ 
unless no column is found at $q = 1$. 
We have observed a similar result for instance ID~15. 
For instance ID~13, 
the set-partitioning constraint set $V^{=}_{(b,\ell)}$ 
has been updated twice 
as it is depicted in the figure. 
We note that $V^{=}_{(b,\ell)}$ has been updated at most once 
for all the other solved instances.

\subsection{Column generation as heuristics}
\mbox{ }

For the unsolved instance IDs~17 and 18 
applied to our branch-and-price framework, 
\mpprobref{mpprob:Pbl} is solved instantly after the timeout. 
Table~\ref{tab:LowerBound} shows lower bounds of the objective value 
obtained in this way. 
The best result for each instance is marked in bold. 
\begin{table}[tb]
\centering
\caption{Lower bound values calculated after timeout of 3,600 seconds.}
\label{tab:LowerBound}
\begin{tabular}{crrrr} \hline
ID & \multicolumn{4}{c}{CG-\mpprobref{mpprob:SS}} \\
   & \multicolumn{1}{c}{SPR}    & \multicolumn{1}{c}{SPR}    
   & \multicolumn{1}{c}{No-SPR} & \multicolumn{1}{c}{No-SPR} \\
   & \multicolumn{1}{c}{MCP}    & \multicolumn{1}{c}{No-MCP} 
   & \multicolumn{1}{c}{MCP}    & \multicolumn{1}{c}{No-MCP} \\ \hline\hline 
17 & $-33.63636$ & $-33.63636$ &   \textbf{6.63063} & \textbf{6.63063} \\
18 &   \textbf{27.05238}  &   13.50693  &   3.57018 & 3.57018 \\ \hline
\end{tabular}
\end{table}
These values have not exceeded 
those obtained by \citet{Santiago.2017}. 
This fact shows that the column generation is not necessarily 
a better heuristic method than the state-of-the-art heuristics 
for instances that cannot be exactly optimized. 

\section{Concluding Remarks}
\label{sec:Conclusions}
This paper has presented an exact algorithm 
for the modularity density maximization 
to provide a clustering solution of an undirected graph. 
The problem can be modeled as the ILP formulation 
of the set-partitioning problem, 
and we have proposed a branch-and-price framework for that. 
The acceleration techniques called 
the set-packing relaxation and the multiple-cutting-planes-at-a-time, 
combined with the newly introduced simpler MILP formulation of 
the column generation subproblem which is solved repeatedly, 
have enabled us to find the new and optimal solutions 
for the famous test instances: 
Graph, Les~Mis\'{e}rables, A00~main and Protein p53. 
Political books as well as Protein p53 instances that have over 100 vertices 
have been optimized in around four minutes by a PC. 
Our solution method is deterministic 
and the computation time is not affected 
by any stochastic behavior. 
For Protein p53 instance, 
column generation at the root node of the branch-and-bound tree 
has provided a fractional upper bound solution 
and our algorithm have found an integral optimal solution after branching, 
which justifies the branch-and-price. 

Future work would include a combination of 
heuristics and our formulations with the acceleration techniques 
to find a column in the column generation phase 
for unsolved instances that have over 110 vertices. 
Such heuristic methods may avoid much computation time 
that is necessary if we try to find a column 
by solving our MILP or possibly any other optimization-based problems, 
for a given particular dual solution value of the restricted master problem. 
Nonetheless we must optimize the column generation subproblem 
expressed by any form at some iteration 
when we decide whether we can terminate the column generation or not. 
It is unclear if the optimization problem at that iteration 
is easy to solve.

\section*{Acknowledgments}
The authors thank Alberto Costa for sharing the instance data files with them.

%
%

\end{document}